# SPECIAL RELATIVITY FOR THE SCHOOL GOING CHILD

T. P. Singh
*Tata Institute of Fundamental Research*

Many years ago, there used to live in Colaba two smart young sisters, Amrita and Rujuta. Amrita was at that time studying in the tenth standard, and Rujuta was in eighth. Both were very interested in science, and their teachers and parents used to tell them wonderful things about the world around them. One day Amrita came home from school, all excited. She said to her sister, ``You know Ruju, today my teacher told us about Einstein's special theory of relativity, and she did such a good job of explaining it, that I think I understand it quite well. You want to know?" Of course I do, said Rujuta. Tell me right now. OK then, lets grab some chips, and get down to work! Its going to take some time, but it will be worth the wait. In the end you will have learnt something fascinating about space and time, and about light, and about how things move, when they move very fast.

## CHAPTER I

## Motion is Relative

When you look at the world around you, you see lots of things. Some may be moving, and some may be at rest. You might see a car moving by at high speed, and you might see a table which is not moving. You are also aware that all these things live in space. And you are also aware of time – time that seems to be passing by. There is yesterday, today and tomorrow. Time that is measured by clocks and watches. The special theory of relativity, which was discovered by Einstein in the year 1905, tells us new things about space and time – things that we simply cannot imagine by looking at our day to day world.

Lets start by talking about motion of things. First, I want to make sure you understand the concept of speed. Speed means how fast something is moving. If the speed of an object does not change with time, we say it is moving at a *uniform* speed. This speed can be calculated by dividing the distance travelled by the time taken to cover that much distance. For example, if one day you decide to walk to your school, which is 3 kms away, and if you take one hour to do this, your speed will be three kilometers per hour.

Please remember one thing. We will only be talking about uniform speeds here. It would be too much to keep on saying uniform speed, uniform speed all the time. But that's the kind of speed we will always be talking about.

Now Rujuta, there is a very important thing you must understand. All speeds are *relative*! What on earth does that mean? Its quite simple actually. Suppose we are both sitting side by side in a bus, and the bus is moving at a uniform speed of sixty kilometers per hour. Then what is Rujuta's speed? Its sixty kilometers per hour. What is Amrita's speed? Also sixty kilometers per hour. Both of us are moving, but are we moving away from each other? No, not at all. We are all the time sitting side by side. So Rujuta is not moving with respect to Amrita. As seen by Amrita, Rujuta is in fact at rest! So we say that the relative speed of Amrita and Rujuta is zero. They are both moving, but with respect to the ground outside, not with respect to each other. Their speed relative to the ground is sixty kilometers per hour, but their speed relative to each other is zero.

To make this clearer, let us imagine that our dear friend Kaustubh is standing on the road and watching the bus go by. We say that our speed relative to Kaustubh is sixty kilometers per hour. We might even imagine that the bus was not moving at all, but Kaustubh, and the ground, and all the other things outside, were rushing past us in the opposite direction, at a speed of sixty kilometers per hour!

So my dear little sister, remember, all speeds are relative. In fact, *all motion is relative*. The bus is moving relative to the ground. We, in the bus, are moving relative to the ground. We are *not* moving relative to each other. Kaustubh is not moving relative to the ground. But he is moving, relative to us.

How can we calculate the relative speed of moving objects? Suppose I am going in a car at fifty kilometers per hour, and Kaustubh is in another car just by the side of my car, also moving in the same direction at fifty kilometers per hour. Then its obvious that our cars are going to remain side by side all the time, and that our relative speed is zero. We got this result by subtracting one speed from the other. This answer is expected, because we are at rest with respect to each other.

Now think of a situation where Kaustubh's car is moving faster than my car, say at seventy kilometers per hour. What is our relative speed now? Is it zero? No, it cannot be. Since he is moving faster, he will overtake me and appear to move away from me. So our relative speed cannot be zero. At what speed will he appear to go away? Seventy kilometers per hour? No. That is his speed with respect to the ground, not with respect to me. His speed relative to me is the difference of our speeds, seventy kilometers per hour minus fifty kilometers per hour, that is, twenty kilometers per hour. To me, he will appear to be moving at twenty kilometers per hour. To him, I will appear to be moving at twenty kilometers per hour, in the opposite direction.

What if Kaustubh's car was actually coming towards my car from the opposite direction, at seventy kilometers per hour? Since we are both moving towards each other, it should be obvious that our relative speed will be higher than seventy, and higher than fifty. It is in fact the sum of our speed, seventy plus fifty, equal to one hundred twenty kilometers per hour.

So you have probably now understood the rule for finding relative speeds of two objects moving in the same direction or in the opposite direction. In the first case, you subtract their speeds, and in the second case you add them, in order to find their relative speed.

Consider another example. This time you, Rujuta, are sitting in a train which is going past a train station on which I am standing. Suppose again that the train is moving at a speed of fifty kilometers per hour. And lo and behold, on the train's roof, a crazy man is running towards the engine (like they sometimes do in the movies)! If he is managing to run towards the engine, then clearly he is not at rest relative to the train. (In fact, Ruju you are at rest relative to the train, because you are sitting lazily on your seat). So he has some speed relative to the train. Let us suppose his speed relative to the train is ten kilometers per hour. Are you clear what that means? It means that if Rujuta, you could see through the roof of the train, you would find him moving forward at ten kilometers per hour.

The question is, what is the the man's speed relative to me? Remember, I am standing on the ground. Is the man's speed relative to the ground ten kilometers per hour, or fifty kiolometers per hour? Neither of these, in fact. Look at it this way. You, Ruju, are moving, say to my left, at fifty kilometers per hour. This would also have been the man's speed, had he not been crazy enough to run on the roof. But in addition to having this speed of fifty kilometers per hour, he also has an extra speed, ten kilometers per hour, relative to the train, and in the same direction as the train. So Ruju, as should be obvious to you, the man is moving to my left faster than you are! His speed relative to me (and relative to the ground) is actually fifty kilometers per hour, *plus* ten kilometers per hour, which makes it sixty kilometers per hour.

Lets summarize this important result. The man's speed relative to the train is ten kilometers per hour. His speed relative to the ground is a *different* number, sixty kilometers per hour, obtained by adding the man's speed relative to the train, and the train's speed relative to the ground.

Rujuta was quite excited with all this, but felt like she needed a break. It was play-time, so the sisters decided to call it off for a while, and went down to play with their friends. At the back of Rujuta's mind, though, there was this man running on the train – hey man, be careful, don't fall off …But what *is* special relativity, she thought? Is it about people running on trains, and about people going past each other in fast moving cars? As she would soon find out, there was more to it, much more. Things that she could never have even dreamt of.

# CHAPTER II

# The Speed of Light

After dinner, the sisters were at it again. And the next day was Saturday, so Mummy was not going to tell them to sleep right away. Huddled in their bedroom, Amrita started once again.

Ruju, I now want to tell you about light. Of course, you know what light is. Like every evening we say, put on the lights. Or, sunlight, it is light from the sun. Or, you switch on a torch, and you see a beam of light coming out from the torch. Next time round, when you have a torch in your hand, go out into the dark night, and point the torch to the sky before switching it on. And when you switch it on, a beam of light will shoot out of the torch, and go out far, far and even farther. It will keep going.

Going? Yes, going. And that means that light travels, and anything that travels must have a speed! So you see, we are back again to talking about speed. What then is the speed of light? Its not a easy thing, to measure the speed of light. But many many years ago, scientists did succeed in finding out how fast light moves. And it moves very, very fast indeed. The speed of light is three lakh kilometers per second. Wow! That's huge. It takes light about one second to reach from the earth to the moon. And while an aeroplane (which I thought moves quite fast!) would take about an hour or two to go from Mumbai to Delhi, a beam of light would cover that distance in much less than one second.

Now, Rujuta, will you please get back on to the train? Thanks. I want to talk about relative speeds again. The relative speed of light. Remember, you were on the train, moving past the station platform, at fifty kilometers per hour. And the man atop the roof, moving at ten kilometers per hour relative to you, and at sixty kilometers per hour relative to the ground.

Now we are going to replace the man by a beam of light. Suppose you are sitting in the train facing towards the engine. Now switch on a torch, and let the beam of light travel straight ahead. So, instead of the man running on the roof, we have this beam of light moving forward, towards the engine, very fast. What is the speed of light as measured by you, Rujuta? Lets accept that you do an experiment, and find the speed of light to be three lakh kilometers per second. So the speed of light relative to you is three lakh kilometers per second.

Now we come to the most important part of our today's dialogue. The most important fact about special relativity, from which so many extraordinary consequences follow. I ask you, Rujuta, what is the speed of light relative to the ground, that is relative to me (Amrita) standing on the station. You would use the same reasoning as you did for the man running on the roof. Add the speed of light relative to the train, and the speed of the train relative to the ground. So you would say that the result is three lakh kilometers per second plus fifty kilometers per hour. The speed of light relative to the ground is three lakh kilometers per second plus fifty kilometers per hour.

Right? NO! WRONG! Completely wrong. Experiments show that the speed of light relative to the ground is the same as its speed relative to the train! Three lakh kilometers per second. The rule of finding relative speed by adding speeds does not apply to light. Light moves at the same speed relative to everyone.

How can that be, you say. How? Prove it, you say, prove it to me. Take it easy dear. Scientists have learnt this by doing experiments. There is nothing to prove here. Its a fact established by experiment, and we have to simply accept it.

You see, we learn facts about the world around us by doing experiments. Then we make rules and theories which explain those facts in a nice and simple way. Such rules are generally referred to as laws. A few hundred years ago, Galileo and Newton discovered the laws of motion of bodies, on the basis of experiments on how things move. These are known as Newton's laws of motion. You have probably studied about them at school.

Newton's laws of motion correctly explain the motion of objects that we see in our day to day life. Moving balls, moving vehicles and moving planets, for example. From these laws it follows that relative speeds are to be found by adding speeds [remember: adding the man's speed relative to the train and the train's speed relative to the ground gave his speed relative to the ground].

Some two hundred years after Newton, scientists succeeded in doing an experiment to find the relative speed of light. And they found that light moves at the same speed relative to everyone. This means that Newton's laws of motion do not apply to light.

So, Newtons's laws must be replaced by new laws of motion. In the year 1905, Einstein discovered these new laws of motion, according to which the speed of light is the same relative to everyone. These new laws came to be known as Einstein's *Special Theory of Relativity.*

Does this mean that Newton was wrong? No, not at all. Newton's laws of motion correctly describe the motion of bodies so long as the speed of the body is much smaller than the speed of light. If a body starts to move very fast, at a speed comparable to the speed of light, then Newton's laws do not correctly describe its motion. But Einstein's laws do. And, very importantly, Einstein's laws agree with those of Newton when the speed of the body is much smaller than the speed of light. We say that Newton's laws of motion are an approximation to Einstein's laws – an approximation that holds good to a great accuracy for bodies moving slowly, compared to light. The motion of light itself can be correctly described only by Einstein's theory, which tells, in particular, that light moves at the same speed relative to everyone.

In Einstein's theory, the rule for finding relative speeds is different from Newton's. Relative speed is not found merely by adding speeds. The rule is such that the speed of light comes out the same for everyone. And moreover, for bodies moving slow compared to light, the rule becomes the same as that of Newton: relative speeds of such bodies are to be found by adding speeds, just as Newton found. Isn't that nice?

[In fact, if you are comfortable with a little bit of mathematics, I can write down for you the rule for adding speeds, as given by Einstein. Suppose $v_b$ is the speed of a body on the train (this is the man on the roof, or the beam of light). Let us denote by v the speed of the train. Let $v_r$ be the speed of the body relative to the ground. Einstein found that $v_r$ is given by the rule

$$v_r = \frac{v_b + v}{1 + \frac{v_b v}{c^2}}$$

In this formula, c denotes the speed of light. You see, when the speed of the body and the speed of the train are much smaller than the speed of light, the term to the right of the `one' in the denominator is very small and can be ignored, and this rule becomes the same as the addition rule of Galileo and Newton! It is in this sense that Newton's laws of motion are an approximation to those of Einstein.

On the other hand, suppose the body in question is a beam of light, so you must set $v_b$=c. You can easily check that then you get $v_r$=c. Isn't that great? The speed of light comes out the same relative to everyone, according to this rule.]

Einstein also found that no body can move faster than light. Things can move only slower than light, or at best as fast as light. If something is moving at the speed of light, it must always move at the speed of light! It cannot slow down. And if something is moving slower than light, it can never move as fast as light.

By now Rujuta was tired and sleepy. So, with their minds full of relativity and what not, Amu and Ruju fell asleep. And no wonder they dreamt of beams of light, of Einstein and Newton.

# CHAPTER III

## Space and Time

No sooner had they woken up on what turned out to be a nice and rainy morning, Amrita began talking again.

Rujuta, the fact that the speed of light is same for everyone leads to dramatic consequences regarding our understanding of space and time. I will tell you about these now. Its not difficult, what I am going to tell you now, but surely its very surprising! And something that we are not used to in our day to day life.

Do me a favour, Ruju. Will you, please? Get on to our good old train, which moves past me at the platform, at the speed of fifty kilometers per hour. And be ready with your torch. You are again to put on the torch so as to release a beam of light. Only this time, please point the torch upwards, towards the ceiling, where there happens to be a mirror, so that the beam will be reflected and come back exactly to where it started from.
This is what you, Rujuta, will see as the path of light.

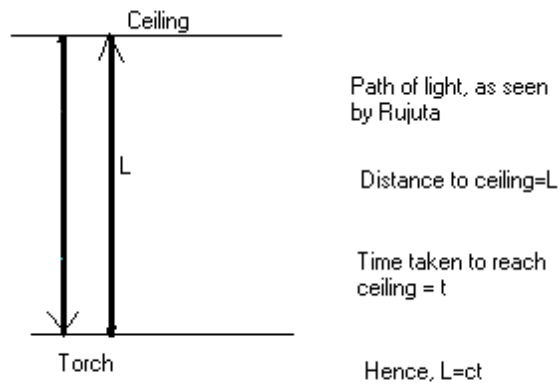

Figure 1

Of course for you the beam will simply go up and down, and the total distance travelled by the beam is twice the distance from the starting point to the ceiling. But what will I, standing on the ground, see as the path of the beam? By the time the beam comes back to the starting point, the train will have moved slightly to the left, so the path of the beam, as seen by me, will *not* be just up and down, but slanting, as shown in the figure below.

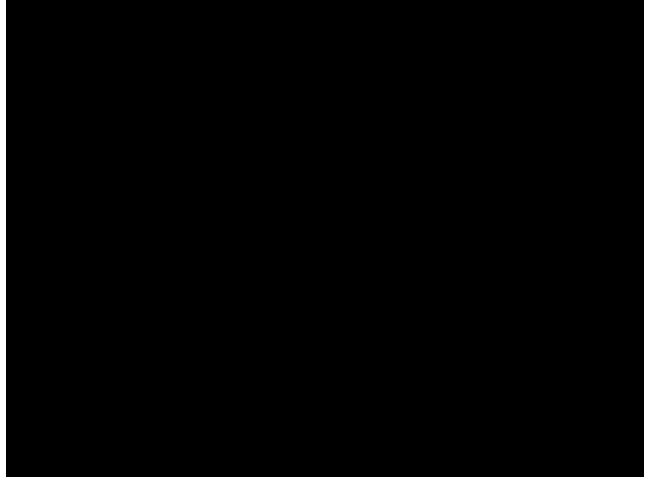

Do Rujuta and Amrita agree that the beam of light travels the same distance, as seen by either of them? No, of course not. Amrita finds that the beam travels more distance than claimed by Ruju. That's because for Ruju the beam simply travels up and down, but for Amu it takes a slanting path. Clearly, for Amu the beam travels a longer distance.

Now the fun can begin. We have agreed a while ago that the speed of light is the same for everyone. The time taken to go up and come down is simply the distance travelled divided by the speed. Speed of light is same for the person on the train, as well as the person on the ground. But distance travelled is different. So to come back to the starting point, light takes more time for Amu, than it does for Ruju!

Lets make that clearer with an example. Both the sisters are wearing a watch each. Suppose when Ruju switches on the torch, both their watches show the time to be ten a.m. When the beam reaches back at the torch after the reflection, let Ruju's watch read five minutes past ten. According to our reasoning above, according to Amu light will take more time to come back to the torch, so Amu's watch will show a larger time, say seven minutes past ten!

[With a little bit of maths, its not hard to calculate the relation between the rates at which the the two clocks move. Look at the two figures above. Suppose according to Ruju light takes a time $2t$ to come back to the torch. This means it will take a time $t$ to reach the ceiling. If $L$ is the distance to the ceiling, we have the relation $L=ct$. According to Amu, let the light beam take a time $2T$ to come back to the torch. So it will take a time $T$ to reach the ceiling, and hence cover the distance $cT$ along the slanting direction. Also, in this much time the train travels a distance $D$ given by
$D = vT$, where $v$ is the speed of the train. By the application of the Pythagoras theorem to the triangle in Figure 2 we have the relation

$$(cT)^2 = (vT)^2 + (ct)^2$$

from which we immediately get, upon simplification, the relation between t and T :

$$T = \frac{t}{\sqrt{1 - \frac{v^2}{c^2}}}$$

Again you see that if the speed of the train is much less than the speed of light, the term after `one' under the square-root can be ignored, as it is very small, and we get T=t, which is the assumption made by Galileo and Newton. We once again see that Newton's mechanics is an approximation to Einstein's mechanics, valid when the speed of the moving bodies is much smaller than the speed of light.]

What is this telling us? If two persons are moving relative to each other, their clocks do not run at the same rate. Amrita, standing on the ground, concludes that Ruju's clock is going slower. This is simply a consequence of light having the same speed for everyone. According to Newton's laws of mechanics, speed of light would not have been same for Amu and Ruju, and in fact, clocks run at the same rate for everyone. [Of course they do, in our daily lives we do not go about quarreling and disagreeing about how fast our clocks are running, even though we move about a lot everyday, relative to each other].

Wait a minute, says Ruju, I am a bit confused. You said the correct laws of motion are those of Einstein, Newton's laws are only approximately true. So why do we not have a problem with our clocks. If I sat on my train, my clock should run slower, shouldn't it? Good point, said Amu, good point. Your clock will indeed run slower, but only by a very very tiny amount. The difference in the rate of your clock and my clock will be very slight, because the train is moving at a speed so much smaller than the speed of light. The difference in the rates of clocks would become easily noticeable if your train starts moving at a speed comparable to the speed of light.

[Ah, said Ruju, I get that one. But there is something else bothering me. You had earlier told me that all motion is relative. So I, Ruju, could as well have imagined that Amrita you are the one who is moving. So I should conclude that my (Ruju's ) clock should be showing more time than Amu's, clock, unlike what you are claiming. Smart, Ruju, smart, said Amrita. You are warming up to the scene. It turns out that both Amu and Ruju are right [funny!] but I am going to leave it to you to sort out that one …]

Something similar happens also for lengths of rods, and in general for the size of things. Suppose my dear sister, you are again travelling in the train [poor you, so much train travel!]. Suppose you have a straight rod with you, placed exactly along the direction in which the train is moving, and you measure the length of the rod, and find it to be some number, say one meter. And me, Amrita, standing on the ground, also measures the length of that rod you are carrying, by doing some experiment. I will find the rod to be less than one meter long! Crazy, but true. And again this is a consequence of the speed of light being the same for everyone.

[If Rujuta measures the length of the rod, placed along the direction of motion of the train, to be L, then according to Einstein, Amrita will measure the length of the rod to be a number D given by the the relation

$$D = L \sqrt{1 - \frac{v^2}{c^2}}$$

Once more, if the speed v of the train is much smaller than the speed of light, the last term under the square root is very small and can be ignored, and we get the expected result of Newtonian mechanics: D=L. In Newton's mechanics different persons moving relative to each other agree on the size of things, just as they agree on the rates at which clocks move.]

Moving objects appear to be shorter than they actually are, according to special relativity, and moving clocks appear to run faster than the actual rate at which they are running. Hence, according to Einstein, there is nothing absolute about space, neither is there anything absolute about time. How much time will have passed, and how much size a body will have, depends on who is making the measurement. People in relative motion do not agree about the results of these measurements. This is completely different from what Newton said. Once again, you might think Newton was wrong. No, he wasn't wrong, he was approximately right. And right to a very good approximation. If the bodies on which we make measurements are moving much slower than light, we will agree that clocks are running at the same rate, and that sizes are same, as seen by people in relative motion.

But Amu, its frustrating. According to Einstein, people in relative motion seem to disagree about everything! I don't like it. I like people to agree with each other, and be friendly. Of course, lil sis, of course. You forgot something! People in relative motion agree about the speed of light. They all agree that light moves at the same speed, three lakh kilometers per second. It is because they agree about speed of light that they are forced to disagree about length and time.

And yet, there is still one more wonderful fact. Something made out of both space and time is there, on which two people in relative motion agree. I will give you an example. Please get back on the train, Rujuta, and this time take Kaustubh with you. Let us imagine that Kaustubh is sitting on the seat opposite you, at a distance one meter away, as measured by *you*. Now throw a ball to him, and suppose it reaches him after one second, again this time is as measured by *you*.

I, Amrita, am standing on the ground, and watching this ball throwing business. Of course, I am not going to agree with you on how far Kaustubh is sitting from you, nor am I going to agree with you on how much time the ball took to reach Kaustubh. That's what we have learnt from Einstein.

But suppose you calculate the following number. Take the square of the time passed, which is one second in your case. Multiply it by the square of the speed of light. From the result subtract the square of the distance between you and Kaustubh, one meter as measured by you. Lets call the resulting number Rujuta's Number.

Now I, Amrita, will do a similar calculation. Having found the time the ball took to reach Kaustubh (as measured by *me* standing on the ground) I will square this time. I will multiply that by the square of the speed of light. From the result I will subtract the square of the distance between Rujuta and Kaustubh (distance as measured by me). Let us call this answer Amrita's Number. Here then is the surprise:

$$\text{Amrita's Number} = \text{Rujuta's Number}$$

I have not proved it to you, but this is another consequence of the speed of light being same relative to everyone.

So, according to Einstein, space is not absolute, and time is not absolute, but something made out of space and time together is absolute. Hence we talk of space and time as being parts of a new concept, spacetime. As if to say that space does not exist without time, and time does not exist without space, both exist in togetherness, as spacetime.

What exactly does this number mean, that we made out of space and time, and on whose value people in relative motion agree? To understand this, we shall introduce the idea of an event. An event is specified by telling when and where something happened. So when you and Kaustubh were exchanging the ball, the first event happened just when you threw the ball. The position of this event was your (Ruju's) position, and the time of this event was the time on your clock when you threw the ball. This is a spacetime event, whose location in space and in time is as I just mentioned.

The second spacetime event happened when Kaustubh received the ball. The position of this event is where Kaustubh is sitting, and the time of this event is the time (according to your clock) when Kaustubh received the ball. The number that we called Rujuta's Number above, measures how `far' the first spacetime event is from the second spacetime event, as seen by Rujuta.

Now on the ground, I (Amrita) also observe these two spacetime events. I also measure the positions of Rujuta and Kaustubh, and the times of the two events, and the number that we called Amrita's number above measures how `far' the two spacetime events are from each other, as seen by Amrita.

The fact that these two numbers are equal means that people in relative motion agree on the `distance' between two spacetime events, even though they do not agree on the amount of time between these two events, or on the length of space between these two events. Remember, this `distance' between spacetime events is not the ordinary distance in space; its distance in spacetime! We also call it the spacetime interval between two events.

As you know Rujuta, space is three dimensional, because there are three directions in space. To this we add the one dimension of time. And we say that spacetime is four dimensional. We live in a four dimensional spacetime. Our Universe consists of matter living in a four dimensional spacetime. Once and for all, the distinction between space and time has been blurred, and this had to happen because the speed of light is the same for everyone!

Once again, the kids needed a break. Also, it was lunchtime. And they still had to take their baths. And do their home-work. And go for a birthday party in the evening. So it was decided that the story of special relativity was to be continued on Sunday morning.

## CHAPTER IV

### Mass and Energy

Newton gave us the laws of motion. The first law says, as I am sure you have studied in school, that unless it is acted on by an external force, a body at rest continues to be at rest, and a body in uniform motion continues to be in uniform motion. This law stays as such, unchanged, in Einstein's special theory of relativity.

Newton's second law says that the state of uniform motion of a body can only be changed by applying a force to the body. In fact you may have also learnt that the force applied to the body is equal to the rate of change of its momentum. Now what is momentum? Well let me give you a simple-minded picture: momentum is what you get by multiplying the mass of a body with its speed. Momentum is an important concept because momentum is what changes when a force is applied to a body. It is a measure of the mass of a body as well as of its speed

In special relativity too, the second law of motion is true, except that here the definition of momentum changes. The momentum still depends on mass and speed, but it depends on speed in a way different from that in Newton's theory. In fact if a body starts moving very, very fast and its speed approaches that of light, its momentum starts to become infinitely large.

[The definition of momentum in special relativity is:

$$p = \frac{mv}{\sqrt{1 - \frac{v^2}{c^2}}}.$$

Here v is the speed of the body, c is the speed of light, and m is the mass of the body. You can see for yourself that as v approaches c, the momentum becomes infinite. This is one way of understanding why nothing can move faster than light. Also, you can see that for speeds small compared to the speed of light, you can ignore the term after one in the square root, and you recover Newton's definition for momentum.]

In Newton's mechanics one also talks of the energy of motion of a body, which we call its kinetic energy. Ruju you do have an idea as to what we mean by the energy of a body, don't you? In Newton's mechanics, the kinetic energy of a body is calculated by multiplying the mass of the body by the square of its speed, and then dividing the result by two. You can easily check that the kinetic energy is also equal to the square of the momentum divided by two times the mass of the body. This is the relation between the energy and the momentum of the body in Newton's mechanics. If the momentum is zero, the energy is also zero.

You can also think of kinetic energy in another way. A force acting on a body does work on it, and this work done on the body is converted into its kinetic energy. In fact the mathematical expression for kinetic energy is what it is, precisely because of the second law, and because momentum is mass times speed.

If you apply force to a body, you change its speed, and hence you also change its kinetic energy. The same is true in special relativity, except that the definition of kinetic energy in terms of speed is again different from that in Newton's mechanics. The most interesting thing is that the relation between energy and momentum is now very different. If you square the energy of the body and subtract from it the square of the momentum multiplied by the square of the speed of light, the resulting number is always equal to the square of the mass multiplied by the fourth power of the speed of light! The speed of light never appears in the energy-momentum relation in Newton's mechanics, but in special relativity it does.

[In special relativity the kinetic energy E of a body is defined as

$$E = \frac{mc^2}{\sqrt{1 - \frac{v^2}{c^2}}}$$

which becomes infinite when the speed of the body approaches the speed of light. Using this expression and the definition of momentum given above we easily see that
$$E^2 - p^2 c^2 = m^2 c^4 \quad ].$$

This relation between the energy and momentum of a body is true from everyone's point of view. Thus if Ruju was in the moving train and measured the energy and momentum of a ball moving in the train, she would find the above relation. And if Amu, standing on the train station, measured the energy and momentum of the body, she would get different answers for the values of energy and momentum, compared to Ruju, but the relation between energy, momentum and mass would be the same as obtained by Ruju.

The real interesting thing is that unlike in Newton's mechanics, the energy of the body does not go to zero even when its speed (and hence momentum) goes to zero. From the above relation, we instead get, for a body at rest, **E = m c$^2$** . This is one of the most famous, if not *the* most famous, equations of physics.

So we learnt for the first time, from Einstein, that there is really no difference between mass and energy! Mass is just a form of energy. If a body at rest has some mass m, and if you want to find out how much energy it has, just multiply its mass by the square of the speed of light. Today, physicists all the time use this relation between mass and energy, essentially with their eyes closed (!), but when Einstein first discovered this relation (again as a consequence of the speed of light being the same for everyone) it was a revolution.

Indeed, this relation between mass and energy has been verified experimentally many, many times. It also plays a very important role in nature, and in the man made world. The relation also tells us that mass and energy can be converted into each other. For example, did you know that the energy of sunlight comes because the sun is converting its mass to energy? Deep inside the sun, hydrogen is converted into helium, and because a helium nucleus has less mass from the four hydrogen nuclei from which it is formed, the lost mass actually shows up as the energy of sunlight!

And did you know that the atom bomb works on the same principle? Conversion of mass into enormous energy. Alas, during the second world war Einstein's ideas were put to misuse, and the atom bombs dropped on Hiroshima and Nagasaki in Japan lead to enormous destruction. But that was not Einstein's fault, you see. We as human beings have to learn to put scientific discoveries to good use. For example, the energy obtained from converting mass is a useful source of energy in our daily lives.

## WHAT NEXT?

With that, dear Ruju, I have come to the end of my story. I have told you most of what my teacher told me on this topic. This is the special theory of relativity, that Einstein discovered in 1905. By the way, I should add that the name `Special Relativity' was not given to the theory in 1905. In fact, the famous research paper of Einstein, in which this discovery was reported, was titled `On the electrodynamics of moving bodies'. Scientists report their results by writing research papers and by publishing them in science journals.

The name `Special Theory of Relativity' was given by Einstein some ten years later, when he discovered another theory, which he called the `General Theory Of Relativity'. In the General Theory, Einstein proposed a new law of gravitation, and modified Newton's law of gravitation. [You would think Einstein was really hell-bent on changing everything that Newton did! But Einstein had the highest respect for Newton, and regarded Newton as the greatest physicist of all times].

You know Ruju, many other great physicists also did important work on this topic, and on the speed of light, *before* Einstein. Some of them were Maxwell, Lorentz, Poincare, Michelson and Morley, and Fitzgerald. We must not forget them. Their contribution was also important. Einstein happened to come on top because his findings were the clearest, and the most general.

In the end I want to tell you the most fascinating thing my teacher told me. She said many physicists today are not satisfied with Einstein's special theory of relativity! They think there are reasons why Einstein's theory should be replaced by another theory, just as there were reasons to replace Newton's mechanics by a new theory. And scientists are working on this problem with lot of excitement and enthusiasm. You know what, I think I am going to be a scientist when I grow up. What about you?